\documentclass[11pt,a4paper,english]{article}
\usepackage[utf8]{inputenc} 
\usepackage[T1]{fontenc}    
\usepackage{babel}
\usepackage{blindtext}
\usepackage{hyperref}       
\usepackage{url}            
\usepackage{booktabs}       
\usepackage{amsfonts}       
\usepackage{nicefrac}       
\usepackage{microtype}      
\usepackage{xcolor}         
\usepackage{amsmath,amssymb}
\usepackage{wrapfig}
\usepackage{stfloats}
\usepackage{graphicx}
\usepackage{placeins}  
\usepackage{float}    
\usepackage{caption}
\usepackage{subcaption}
\usepackage{bm}
\usepackage{listings}
\usepackage{multicol}
\usepackage{multirow} 
\usepackage{pgffor}
\usepackage{cleveref}
\usepackage{geometry}

\NewDocumentCommand{\codeword}{v}{%
\texttt{\textcolor{blue}{#1}}%
}

\title{VWAP Execution with Signature-Enhanced Transformers: A Multi-Asset Learning Approach}

\author{%
    Rémi Genet \\
    \small DRM, Université Paris Dauphine - PSL \\
    \small Aplo \\
    \small remi.genet@dauphine.psl.eu \\
}

\begin{document}
\maketitle


\begin{abstract}
In this paper I propose a novel approach to Volume Weighted Average Price (VWAP) execution that addresses two key practical challenges: the need for asset-specific model training and the capture of complex temporal dependencies. Building upon my recent work in dynamic VWAP execution \cite{genet2025dynamicvwap}, I demonstrate that a single neural network trained across multiple assets can achieve performance comparable to or better than traditional asset-specific models. The proposed architecture combines a transformer-based design inspired by \cite{genet2024tkat} with path signatures for capturing geometric features of price-volume trajectories, as in \cite{inzirillo2024sigkan}. The empirical analysis, conducted on hourly cryptocurrency trading data from 80 trading pairs, shows that the globally-fitted model with signature features (GFT-Sig) achieves superior performance in both absolute and quadratic VWAP loss metrics compared to asset-specific approaches. Notably, these improvements persist for out-of-sample assets, demonstrating the model's ability to generalize across different market conditions. The results suggest that combining global parameter sharing with signature-based feature extraction provides a scalable and robust approach to VWAP execution, offering significant practical advantages over traditional asset-specific implementations.
\end{abstract}

\newpage

\section{Introduction}

The execution of large trading orders in financial markets is a complex and consequential challenge, with the potential to significantly impact transaction costs and market dynamics. In this context, the concept of Volume Weighted Average Price (VWAP) has emerged as a key benchmark and execution strategy, offering market participants a robust framework for minimizing market impact while closely tracking average traded prices over a specified period \cite{Madhavan2002}. 
At its core, VWAP execution aims to address two fundamental objectives. First, by distributing an order over time and closely tracking the average traded price, VWAP strategies seek to minimize the market impact of large trades—a key component of overall transaction costs, as established by Berkowitz et al. \cite{TotalCostOfTransactions}. Second, by targeting a pre-defined benchmark, VWAP provides a transparent and objective measure of execution quality, which is crucial for institutional investors.

\subsection{VWAP Definition and Discretization}

Following the seminal work of Konishi \cite{Konishi}, the Volume Weighted Average Price (VWAP) over a time period \([0,T]\) is defined mathematically as:
\begin{equation}
    \text{VWAP}_{[0,T]} = \frac{\int_0^T P(t)V(t)\,dt}{\int_0^T V(t)\,dt},
\end{equation}
where \(P(t)\) and \(V(t)\) represent the price and volume at time \(t\), respectively. As noted by McCulloch and Kazakov \cite{Culoch2007}, financial markets operate in discrete time intervals, leading to a discretized form:
\begin{equation}
    \text{VWAP}_{[0,T]} = \frac{\sum_{t=1}^{T} P_t\,V_t}{\sum_{t=1}^{T} V_t},
\end{equation}
where \(P_t\) and \(V_t\) are the price and volume in the \(t\)-th time interval. For a trader executing a large order of total size \(Q\), Humphery-Jenner \cite{Humphery} demonstrates that the objective is to minimize the difference between the achieved execution price and the market VWAP. Let \(q_t\) denote the quantity traded in interval \(t\), such that:
\begin{equation}
\sum_{t=1}^{T} q_t = Q.
\end{equation}
The execution price achieved is given by:
\begin{equation}
P_{\text{exec}} = \frac{\sum_{t=1}^{T} P_t\,q_t}{Q}.
\end{equation}
Following Bialkowski et al. \cite{LeFol2006}, the VWAP execution problem can be formulated as minimizing the slippage:
\begin{equation}
\min_{q_1,\ldots,q_T} \left|\frac{\sum_{t=1}^{T} P_t\,q_t}{Q} - \frac{\sum_{t=1}^{T} P_t\,V_t}{\sum_{t=1}^{T} V_t}\right|.
\end{equation}
For clarity, the normalized order allocation \(\tilde{q}_t = \frac{q_t}{Q}\) (so that \(\sum_{t=1}^{T} \tilde{q}_t = 1\)) and the normalized market volume profile \(\tilde{V}_t = \frac{V_t}{\sum_{t=1}^{T} V_t}\) (with \(\sum_{t=1}^{T} \tilde{V}_t = 1\)) are introduced. With these definitions, the execution price becomes:
\begin{equation}
P_{\text{exec}} = \sum_{t=1}^{T} P_t\,\tilde{q}_t,
\end{equation}
and the market VWAP is:
\begin{equation}
\text{VWAP} = \sum_{t=1}^{T} P_t\,\tilde{V}_t.
\end{equation}
Genet \cite{genet2025staticvwap} reformulates the slippage as a bound:\begin{equation}
    S_T \le \sum_{t=1}^{T} \left| \bigl(P_t-\text{VWAP}_t\bigr)\tilde{q}_t \right| + \sum_{t=1}^{T} \left| \text{VWAP}_t\Bigl(\tilde{q}_t-\tilde{V}_t\Bigr) \right|,
\end{equation}
where \(\text{VWAP}_t\) denotes the market VWAP computed over interval \(t\). The first term quantifies the impact of price deviations weighted by the trader's participation rate, while the second term captures the error due to discrepancies between the trader's normalized allocation \(\tilde{q}_t\) and the market's volume fraction \(\tilde{V}_t\). Under the volume conservation constraint and non-negativity constraints \(\tilde{q}_t \geq 0\), this decomposition separates the overall slippage \(S_T\) into a price deviation component and a volume allocation error component. This optimization problem is particularly challenging because future prices and volumes are unknown at execution time, which necessitates accurate predictions of market dynamics while managing execution risk \cite{frei}. Furthermore, the increasing sophistication of market participants and the rising prominence of transaction cost analysis (TCA) in institutional trading \cite{Madhavan2002} have amplified the importance of VWAP execution strategies. As markets evolve, executing large orders while minimizing market impact becomes increasingly complex, thereby requiring more advanced approaches to VWAP execution.

\subsection{Classical VWAP Approaches}

The theoretical foundation of VWAP execution strategies emerged from seminal works that established the mathematical framework for optimal order execution. Konishi \cite{Konishi} provided one of the first comprehensive analyses of VWAP strategies, demonstrating that in markets where volume and volatility are uncorrelated, the optimal execution curve mirrors the expected relative market volume curve. He further extended his analysis to cover cases where volume and volatility are correlated, providing mathematical validation for empirical observations and establishing a theoretical benchmark for subsequent research. Building on this foundation, McCulloch and Kazakov \cite{Culoch2007} developed a more sophisticated model incorporating practical constraints and information asymmetries by introducing constrained trading rates and potential information advantages. Their work revealed important stylized facts about expected relative volume patterns—most notably, the characteristic S-shape observed in equity markets and the observation that higher-turnover stocks exhibit less variation in their expected relative volume. Numerous studies, including those by Easley and O'Hara \cite{Easley1987}, Viswanathan and Foster \cite{Foster}, Tauchen and Pitts \cite{Tauchen1983}, and Karpoff \cite{Karpoff1987}, examined the relationship between volumes and other market variables, though primarily at low frequencies. Gourieroux et al. \cite{Gourieroux} contributed significantly to understanding market trading activity, and McCulloch and Kazakov \cite{Culoch2012} extended this work by transforming Konishi's fixed model into a continuous dynamic framework, establishing the crucial connection between optimal VWAP strategies and accurate intraday volume estimation.

\medskip

The evolution of these classical approaches reflects a growing recognition of the complexity inherent in VWAP execution. While these models provided valuable insights and theoretical foundations, they also revealed the limitations of purely static approaches in capturing the dynamic nature of modern markets. This recognition would eventually lead to the development of more sophisticated dynamic approaches and, ultimately, to the application of machine learning techniques in VWAP execution strategies.

\subsection{Dynamic Volume Approaches}

A significant paradigm shift in VWAP execution strategies occurred with the introduction of dynamic volume estimation approaches. Bialkowski et al. \cite{LeFol2006} pioneered this advancement by proposing a novel method for estimating intraday volumes through component decomposition. Their work, later refined in Bialkowski et al. \cite{LeFol2012}, separated volume patterns into two distinct components: one reflecting broader market evolution and another capturing stock-specific patterns. This decomposition enabled more accurate volume predictions by modeling the dynamic component using ARMA and SETAR models, demonstrating substantially improved accuracy compared to traditional static approaches. However, transitioning from simplistic volume modeling to these more advanced methods comes at a cost: such approaches no longer explicitly account for the volume–volatility relationship, as it becomes much more challenging to realistically incorporate both components simultaneously. The shift from static to dynamic approaches was further advanced by Humphery-Jenner \cite{Humphery}, who introduced the concept of Dynamic VWAP (DVWAP) in contrast to the traditional Historical VWAP (HVWAP). Their research highlighted a crucial limitation of historical approaches—their inability to incorporate real-time market information during execution. By developing a framework that adapts to incoming news and market developments, they demonstrated significant improvements over historical methods in both basic VWAP tracking and the management of market dynamics.

\medskip

Alternative theoretical perspectives emerged through the work of Bouchard and Dang \cite{bouchard} and Frei and Westray \cite{frei}, who approached VWAP execution through the lens of stochastic analysis. As Frei and Westray \cite{frei} noted, their derived optimal trading rates depended primarily on volume curves rather than price processes, reflecting the assumption of uncorrelated Brownian motion in price movements. This theoretical framework provided valuable insights into the relationship between volume patterns and execution strategy, even as it highlighted the limitations of purely stochastic approaches. A significant contribution to the practical aspects of VWAP execution came from Carmona and Li \cite{Tianhui}, who examined the strategic considerations at both macro and micro scales. Their research was particularly notable for addressing the practical dilemma faced by brokers in choosing between aggressive and passive orders at the high-frequency level, bringing theoretical insights to bear on practical execution decisions. Guéant and Royer \cite{Gueant} made two crucial contributions that addressed previously understudied aspects of VWAP execution. First, they incorporated a comprehensive market impact model that considered both temporary and permanent effects, addressing a critical concern for institutional investors using VWAP orders to manage large positions. Second, they developed a framework for pricing guaranteed VWAP services using CARA utility functions and indifference pricing. This work represented a significant shift from traditional approaches focused solely on benchmark tracking, introducing a more nuanced understanding of risk-adjusted optimal execution.

\medskip

These dynamic approaches collectively highlighted a crucial insight: while modeling market volumes is important, the assumption of independence between prices and volumes often fails to reflect market reality. This recognition, combined with the increasing availability of computational power and market data, set the stage for the application of more sophisticated analytical techniques, particularly in the domain of machine learning and artificial intelligence.

\subsection{The Rise of Deep Learning in Financial Time Series}

In parallel with these theoretical advances, the field of machine learning has witnessed a rapid proliferation of powerful techniques and architectures, particularly in the domain of deep learning. The field of time series analysis and prediction has been fundamentally transformed by developments in deep learning, particularly in the domain of neural networks. As documented by Sezer et al. \cite{sezer2020financial} in their comprehensive review, deep learning models have increasingly outperformed traditional machine learning approaches across various financial forecasting tasks. The evolution of deep learning architectures for financial applications has been marked by several key innovations. The introduction of Long Short-Term Memory (LSTM) networks by Hochreiter and Schmidhuber \cite{hochreiter1997} addressed the vanishing gradient problem that had limited traditional recurrent neural networks, enabling effective learning of long-term dependencies in sequential data. This was followed by the development of Gated Recurrent Units (GRU) by Cho et al. \cite{cho2014}, offering comparable performance with a more streamlined architecture. A revolutionary step forward came with the introduction of attention mechanisms Bahdanau et al. \cite{bahdanau2014neural}, culminating in the Transformer architecture Vaswani et al. \cite{vaswani2017attention}. While initially developed for natural language processing, these architectures' ability to capture both local and global dependencies in sequential data made them particularly suitable for financial time series analysis.

\medskip

Recent years have seen an explosion of deep learning applications in finance, with researchers tackling increasingly complex challenges. Ackerer et al. \cite{ackerer2020deep} demonstrated the power of neural networks in fitting and predicting implied volatility surfaces, while Horvath et al. \cite{horvath2019deep} showed how deep learning could revolutionize pricing and calibration in volatility models. As highlighted by Zhang et al. \cite{zhang2023deep} in their recent review, deep learning models are gradually replacing traditional statistical and machine learning models as the preferred choice for price forecasting tasks. 

\medskip

In the specific domain of trading volume prediction, significant advances have been made through the development of specialized architectures such as Temporal Kolmogorov-Arnold Networks (TKAN) \cite{genet2024tkan}, Signature-Weighted Kolmogorov-Arnold Networks (SigKAN) \cite{inzirillo2024sigkan}, Temporal Kolmogorov-Arnold Transformers (TKAT) \cite{genet2024tkat}, Kolmogorov-Arnold Mixture of Experts (KAMoE) \cite{inzirillo2024kamoe} and Recurrent Neural Networks with Signature-Based Gating Mechanisms (SigGate) \cite{genet2025siggate}. 

\subsection{Deep Learning Approaches to Market Execution}

The application of deep learning to market execution problems has evolved significantly in recent years. Early approaches focused primarily on using neural networks for price prediction or simple trading signals. However, the complexity of VWAP execution, with its intricate relationship between volume patterns, price impact, and timing decisions, presents unique challenges that require more sophisticated approaches.

\medskip

Recent research has begun to explore more advanced applications of deep learning to execution problems. Papanicolaou et al. \cite{papanicolaou2023optimal} demonstrated the effectiveness of using LSTMs for large order execution within the Almgren and Chriss framework, showing how deep learning models could capture cross-sectional relationships between different stocks' execution characteristics. A significant advancement was achieved with my introduction of Static Neural VWAP \cite{genet2024staticvwap}, which is based on the simple linear internal model described in \cite{genet2024tln}. This approach established a novel method for VWAP execution by leveraging deep learning techniques in a fundamentally different way from existing methods. Instead of focusing on volume curve prediction, as traditional approaches do, I demonstrated that directly optimizing the execution strategy through neural networks can significantly improve performance. Building upon this foundation, I subsequently developed Dynamic Neural VWAP \cite{genet2025dynamicvwap}, which incorporated adaptive capabilities through recurrent neural networks while maintaining the robust performance characteristics of the static approach. This development aligns with earlier findings from Bialkowski et al. \cite{LeFol2012} and Humphery-Jenner \cite{Humphery} about the importance of adapting to changing market conditions. However, like most existing approaches, it required asset-specific training, limiting its practical applicability in markets with numerous assets.

\subsection{Path Signatures}
\label{sec:signatures}

Path signatures, first introduced by Chen \cite{chen1958integration}, provide a powerful way to represent sequential data by capturing its geometric properties through iterated integrals. For a comprehensive treatment of path signatures as a representation for unparameterized paths, I direct readers to \cite{lyons2014rough,lyons1998differential,chevyrev2016primer}. These works demonstrate how signatures can robustly characterize time series by encoding nonlinear temporal dependencies.

\medskip

Recent studies in finance have shown the benefits of incorporating signature features into neural architectures for tasks such as forecasting and clustering. Dyer and Xu \cite{dyer2021deep} demonstrated improved accuracy by integrating path signatures into LSTMs, while Fermanian \cite{fermanian2021embedding} used signature embeddings for unsupervised learning on time series data. Additionally, SigKAN \cite{inzirillo2024sigkan} leverages path signatures to enhance modeling capacity in financial forecasting tasks.

\medskip

In this approach, I incorporate path signatures as additional contextual information within the neural VWAP execution model. This design choice allows us to retain the geometric insights provided by path signatures—thereby facilitating more robust adaptation to varying market regimes—while avoiding extra architectural complexities. By conditioning on signature-derived features that summarize recent price-volume trajectories in a compact form, the model is better equipped to capture nuanced temporal dependencies relevant to VWAP execution.

\medskip

\subsection{The Promise and Limitations of Attention Mechanisms}

The evolution of attention mechanisms, particularly in the context of time series analysis, has opened new possibilities for VWAP execution while also revealing important limitations. While the Transformer architecture \cite{vaswani2017attention} revolutionized sequence modeling in natural language processing, its application to financial time series has proven more nuanced. The self-attention mechanism's ability to capture long-range dependencies makes it potentially valuable for identifying complex market patterns, yet the inherently different nature of financial data compared to language poses unique challenges. Recent studies have questioned the universal applicability of transformers across different domains. As demonstrated by Zeng et al. \cite{zeng2022transformers}, while transformers excel at extracting semantic correlations in language tasks, their effectiveness can be limited when applied to numerical time series where temporal ordering is paramount. This has led to various attempts at adapting transformer architectures for time series, including LogTrans \cite{li2019enhancing}, Informer \cite{zhou2021informer}, and Autoformer \cite{xu2021autoformer}, each introducing specialized mechanisms to handle temporal data more effectively. Nevertheless, specialized transformer architectures have shown promise in specific financial contexts. The Temporal Fusion Transformer (TFT) \cite{lim2021temporal} introduced innovations specifically designed for temporal data, including variable selection networks and temporal attention mechanisms that better preserve time-based relationships. Further developments like Pyraformer \cite{liu2021pyraformer}, which employs pyramidal attention patterns, and FEDformer \cite{zhou2022fedformer}, which incorporates frequency domain analysis, have demonstrated how architectural innovations can enhance transformer performance on time series tasks.

\medskip

More recently, the Temporal Kolmogorov-Arnold Transformer (TKAT) \cite{genet2024tkat} demonstrated significant improvements in volume prediction tasks by combining transformer architectures with the theoretical foundations of Kolmogorov-Arnold networks \cite{kolmogorov1961representation}. TKAT's success in volume prediction, a task closely related to VWAP execution, suggests that carefully designed transformer architectures can be valuable when adapted to specific financial applications. A key consideration in applying transformers to VWAP execution is the role of causal masking. Unlike language tasks where bidirectional context is valuable, financial execution requires strict causality to prevent look-ahead bias. This necessitates careful design of attention masks to ensure predictions depend only on available information, as highlighted by Li et al. \cite{li2019enhancing}. The masking strategy becomes particularly crucial in long-term forecasting scenarios where maintaining temporal coherence is essential for execution quality. The effectiveness of transformers in financial applications appears to be highly dependent on architectural choices that address the specific challenges of market data. While the universal approximation capabilities of transformer networks make them theoretically powerful, their practical utility in VWAP execution requires careful consideration of temporal preservation, causality constraints, and the specific characteristics of market volume patterns.

\subsection{Proposed Innovations}
\label{sec:innovations}

This paper builds upon the parallel developments in neural VWAP execution and attention mechanisms to address two fundamental challenges: the need for asset-specific training and the limitations of traditional architectures in capturing complex temporal dependencies. I propose a framework that combines a dynamic VWAP architecture with signature-based contextual features and enhanced temporal modeling capabilities.

\medskip

First, I demonstrate that a single neural network trained on multiple assets can achieve comparable or superior performance to asset-specific models, substantially reducing the operational complexity of deployment without sacrificing execution quality. Inspired by the signature-based methods in \cite{inzirillo2024sigkan}, I incorporate learnable weights into the signature computation process. Rather than using signatures to weight different experts, this approach treats signatures purely as additional contextual features. This retains the geometric insights provided by path signatures while streamlining the execution model into a single network. Second, I replace traditional recurrent networks with a modified TKAT architecture \cite{genet2024tkat}, leveraging causal attention masks to ensure forward-looking-free operation. This design is crucial for real-time execution and aligns with the practical constraints of VWAP trading. Additionally, the integration of variable selection networks helps the model better capture complex interactions among market variables, while maintaining computational efficiency suitable for intraday trading scenarios. To evaluate the proposed framework, I perform an extensive empirical analysis using hourly trading data from 80 cryptocurrency pairs on the Binance exchange, with data extending to July 1, 2024. The assets are ranked by liquidity and split into training and testing sets by taking every other asset, resulting in 40 training assets. I compare the following four model configurations:

\begin{enumerate}
    \item \textbf{Asset-Fitted DynamicVWAP (AFD):} An asset-specific dynamic VWAP model, following the methodology in \cite{genet2025dynamicvwap}.
    \item \textbf{Globally-Fitted DynamicVWAP (GFD):} A dynamic VWAP model trained jointly on all assets (rather than one asset at a time).
    \item \textbf{Globally-Fitted Dynamic Transformer (GFT) without Signature:} An ablation version of the Transformer-based approach, excluding path signature features.
    \item \textbf{Globally-Fitted Dynamic Transformer (GFT-Sig) with Signature:} The full Transformer-based approach incorporating learnable path signature features as additional context.
\end{enumerate}

\noindent This experimental design allows us to test both the benefits of global training across multiple assets and the added value of signature-based contextual features. Furthermore, by including an ablation version of the Transformer (without signatures), I can isolate the performance gains attributed to path signatures and demonstrate their efficacy in VWAP execution tasks.

\newpage

\section{Signature-Based Dynamic VWAP Architecture}

The proposed architecture builds upon the signature-enhanced dynamic VWAP framework described in \cite{genet2025dynamicvwap} and introduces two key innovations: the incorporation of a transformer-based architecture inspired by the Temporal Kolmogorov-Arnold Transformer \cite{genet2024tkat} and the introduction of learnable signature weights for additional contextual information as in SigKAN \cite{inzirillo2024sigkan}. These enhancements enable more sophisticated temporal pattern recognition while maintaining the framework's ability to generalize across different assets and market conditions.

\subsection{Model Overview}

The architecture processes input sequences at two different temporal scales:
\begin{equation}
    X_{t-l_s+1:t} \in \mathbb{R}^{l_s \times d} \quad\text{and}\quad X_{t-l+1:t+h-1} \in \mathbb{R}^{(l+h-1) \times d},
\end{equation}
where $l_s$ denotes the signature lookback length for capturing long-term geometric patterns, $l$ represents the transformer lookback length for local temporal dependencies, $h$ is the prediction horizon, and $d$ is the feature dimension.

\begin{figure}[H]
    \centering
    \includegraphics[width=0.75\linewidth]{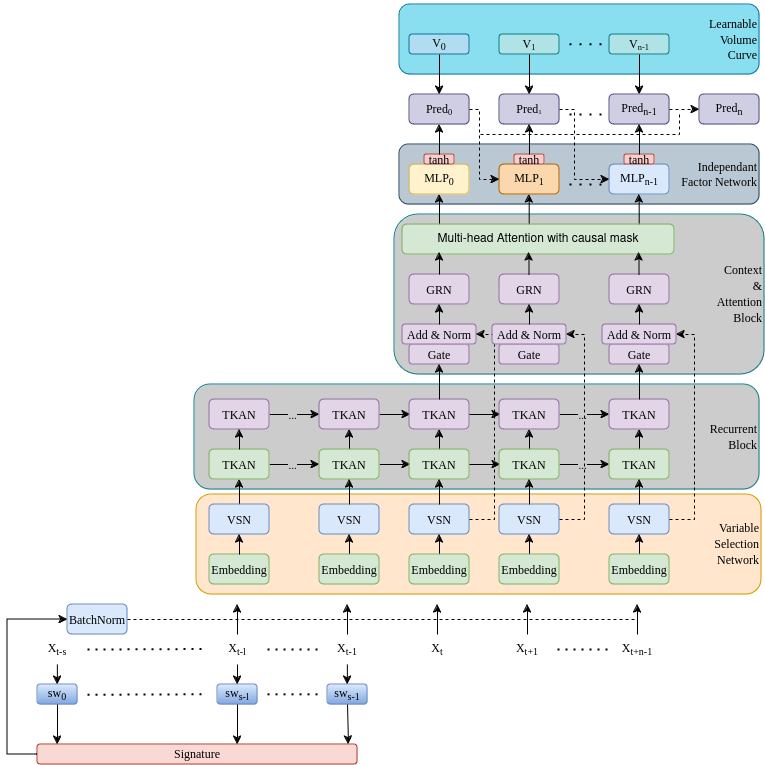}
    \caption{Signature-Based Dynamic VWAP Architecture}
    \label{fig:model_overview}
\end{figure}

\subsection{Transformer-Based Architecture}

The model implements a sophisticated temporal processing pipeline that combines feature selection, temporal modeling, and volume prediction capabilities in a single end-to-end framework. The processing consists of several sequential layers, each serving a distinct purpose in capturing market dynamics:

\begin{equation}
    e_t = \text{Embedding}(x_t)
\end{equation}
The embedded features first undergo adaptive feature selection through a Variable Selection Network (VSN):
\begin{equation}
    s_t = \text{VSN}(e_t).
\end{equation}
A stack of TKAN (or transformer-like) layers then processes these selected features to capture temporal dependencies:
\begin{equation}
    h_t = \text{TKAN}_L\bigl(\dots\,\text{TKAN}_1(s_t)\bigr),
\end{equation}
The TKAN outputs are combined with the original selected features through a gated residual connection:
\begin{equation}
    c_t = \text{LayerNorm}\bigl(s_t + \text{Gate}(h_t)\bigr),
\end{equation}
These combined features undergo further enrichment through a Gated Residual Network (GRN):
\begin{equation}
    r_t = \text{GRN}(c_t).
\end{equation}
Finally, causal self-attention is applied to capture long-range dependencies while maintaining temporal causality:
\begin{equation}
    a_t = \text{MultiHead}(r_t, r_t, r_t, \text{mask} = \text{causal})
\end{equation}

\subsection{Volume Prediction}

Given the processed temporal context, the model generates volume predictions through a sequential process that ensures conservation across the trading horizon. At each time step $t$, the model accesses the temporal context:
\begin{equation}
   h_t = a_{t+\text{lookback}}.
\end{equation}
After the initial allocation, subsequent predictions incorporate previous volume decisions:
\begin{equation}
   h_t = 
   \begin{cases}
       a_{t+\text{lookback}} & \text{if } t = 0 \\
       [\,a_{t+\text{lookback}}\,;\,v_{1:t-1}] & \text{otherwise}
   \end{cases}
\end{equation}
The model computes an adjustment factor to a learned base volume curve:
\begin{equation}
   \alpha_t = 1 + \tanh\bigl(f_t(h_t)\bigr),
\end{equation}
where $f_t$ is a feed-forward network:
\begin{equation}
   f_t(h) = W_3\,\text{ReLU}\bigl(W_2\,\text{ReLU}(W_1h + b_1) + b_2\bigr) + b_3.
\end{equation}
The volume allocation enforces conservation through clipping:
\begin{equation}
   v_t = \text{clip}\bigl(\alpha_t \,v_b^{(t)},\; 0,\; 1 - \sum_{j=1}^{t-1} v_j\bigr).
\end{equation}
The final time step allocates any remaining volume:
\begin{equation}
   v_T = 1 - \sum_{t=1}^{T-1} v_t
\end{equation}

\subsection{Variable Selection Networks}

The Variable Selection Network (VSN) serves as the initial processing layer in the architecture, enabling the model to adaptively focus on the most relevant input features for VWAP execution. This component is particularly important in financial time series where different variables may have varying importance depending on market conditions. The VSN becomes especially valuable when dealing with a large number of input variables, as it is well-known that including too many inputs can greatly impact a model's quality and performance. By adaptively selecting the most relevant features, the VSN helps mitigate this issue. The inclusion of the VSN is particularly beneficial in this model as it incorporate path signatures, which can potentially introduce irrelevant terms. The VSN effectively filters out these terms, ensuring that only the most informative signature components are used. Moreover, this sets the stage for future research to easily extend the model with additional covariates without the risk of deteriorating performance due to irrelevant inputs.
\begin{figure}[H]
    \centering
    \includegraphics[width=0.8\linewidth]{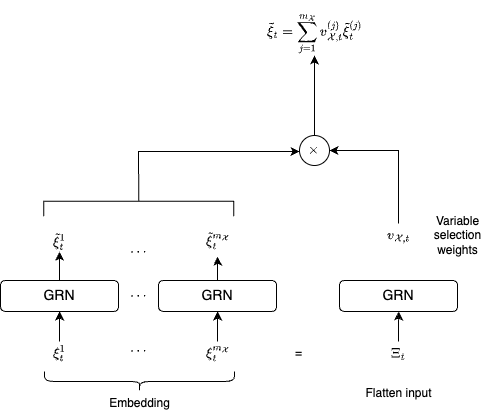}
    \caption{Variable Selection Network Architecture}
    \label{fig:vsn}
\end{figure}
Given an input tensor $X_t \in \mathbb{R}^{B \times T \times E \times V}$, where $B$ is the batch size, $T$ is the number of time steps, $E$ is the embedding dimension, and $V$ is the number of variables, the VSN processes this input through several stages of transformation and selection.

\subsubsection{Initial Feature Embedding}

Each input variable undergoes independent processing through a dedicated embedding layer:
\begin{equation}
    \xi_t^{(j)} = \text{Dense}_j(X_t[:,:,j]) \in \mathbb{R}^{B \times T \times E},
\end{equation}
where separate dense layers are maintained for each variable $j \in \{1,\dots,V\}$, allowing for variable-specific transformations. This initial embedding captures the unique characteristics of each feature while maintaining their temporal relationships.

\subsubsection{Feature Importance Scoring}
The embedded features are then flattened and processed through a Gated Residual Network (GRN) to compute variable importance scores:
\begin{equation}
    \Xi_t = [\xi_t^{(1)^T}, \dots, \xi_t^{(V)^T}] \in \mathbb{R}^{B \times T \times (E \cdot V)},
\end{equation}
\begin{equation}
    v_{\chi_t} = \text{Softmax}\bigl(\text{GRN}_{v_\chi}(\Xi_t)\bigr) \in \mathbb{R}^{B \times T \times V}.
\end{equation}
This importance scoring mechanism enables the model to dynamically adjust its attention to different input variables based on their relevance to the current market state.

\subsubsection{Non-linear Feature Processing}
Concurrently with importance scoring, each embedded variable undergoes non-linear processing through its own GRN:
\begin{equation}
    \tilde{\xi}_t^{(j)} = \text{GRN}_{\tilde{\xi}(j)}(\xi_t^{(j)}) \in \mathbb{R}^{B \times T \times E}.
\end{equation}
This parallel processing allows the network to capture complex patterns within each variable independently while maintaining weight sharing across time steps.

\subsubsection{Weighted Feature Combination}
The final stage combines the processed features using the computed importance weights:
\begin{equation}
    \tilde{\xi}_t = \sum_{j=1}^V v_{\chi_t}^{(j)} \,\tilde{\xi}_t^{(j)} \;\in \mathbb{R}^{B \times T \times E}.
\end{equation}
This weighted combination ensures that the model can dynamically adjust its focus on different input variables based on their predicted relevance to the VWAP execution task. The resulting output preserves the most informative aspects of each feature while suppressing less relevant signals.
\begin{enumerate}
    \item It reduces the impact of noise by automatically "downweighting" less relevant features,
    \item It helps the model focus on the most predictive signals for different market regimes,
    \item It provides interpretable importance weights that can be monitored for stability and reasonableness,
    \item It enables the model to adapt its feature utilization as market conditions evolve.
\end{enumerate}

\subsection{Gated Residual Networks}

The relationships between dependent random vectors is a key issue in financial time series analysis. Gated residual networks offer an efficient and flexible way of modeling complex relationships in time series, as demonstrated by \cite{lim2021temporal}. They allow control over the information flow and facilitate learning tasks, making them particularly useful in areas where non-linear interactions and long-term dependencies are crucial. The implementation follows the architecture proposed by \cite{lim2021temporal}, with modifications to remove context dependencies.

\begin{figure}[H]
    \centering
    \includegraphics[width=0.4\linewidth]{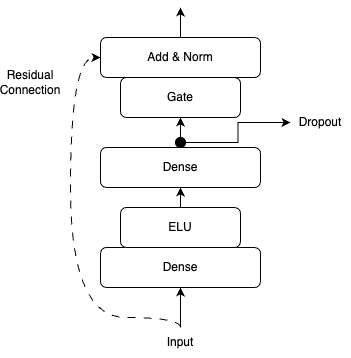}
    \caption{Gated Residual Network Architecture}
    \label{fig:grn}
\end{figure}

\subsubsection{Network Structure}

The GRN processes input features through a series of transformations, combining skip connections with gating mechanisms. Given an input $x \in \mathbb{R}^{d_{model}}$, the network computes:
\begin{align}
    \text{GRN}_\omega(x) &= \text{LayerNorm}\bigl(x \;+\; \text{GLU}_\omega(\eta_1)\bigr) ,\\
    \eta_1 &= W_{1,\omega}\,\eta_2 \;+\; b_{1,\omega}, \\
    \eta_2 &= \text{ELU}\bigl(W_{2,\omega}\,x \;+\; b_{2,\omega}\bigr).
\end{align}
Here, ELU (Exponential Linear Unit) serves as the activation function \cite{clevert2015fast}, providing smooth gradients for negative inputs. The intermediate layers $\eta_1, \eta_2 \in \mathbb{R}^{d_{model}}$ represent the network's internal representations, and $\omega$ is an index used to indicate weight sharing across components. The standard layer normalization $\text{LayerNorm}$ follows the implementation described in \cite{ba2016layer}.
\begin{equation}
    \text{ELU}(x) = 
    \begin{cases}
        x & \text{if } x > 0 \\
        \alpha\,(e^x - 1) & \text{if } x \le 0
    \end{cases}
\end{equation}
\subsubsection{Gating Mechanism}

A Gated Linear Units (GLUs) is implemented \cite{dauphin2017language} to provide adaptive control over the network's non-linear contributions. GLUs offer the flexibility to suppress any parts of the architecture that are not required for a given input pattern. For an input $\gamma \in \mathbb{R}^{d_{model}}$, the GLU computes:
\begin{equation}
    \text{GLU}_\omega(\gamma) \;=\; \sigma\bigl(W_{4,\omega}\gamma + b_{4,\omega}\bigr) \;\odot\; 
    \bigl(W_{5,\omega}\gamma + b_{5,\omega}\bigr),
\end{equation}
where $\sigma(\cdot)$ represents the sigmoid activation function, $W_{(\cdot)} \in \mathbb{R}^{d_{model}\times d_{model}}$ and $b_{(\cdot)} \in \mathbb{R}^{d_{model}}$ are weights and biases respectively, and $\odot$ denotes the Hadamard (element-wise) product.

\subsubsection{Layer Normalization}

The final layer normalization ensures stable training and helps manage the scale of features:
\begin{equation}
    \text{LayerNorm}(x) \;=\; \gamma \;\odot\; \frac{x - \mu}{\sqrt{\sigma^2 + \epsilon}} \;+\; \beta,
\end{equation}
where $\mu$ and $\sigma^2$ are the mean and variance computed over the feature dimension, $\gamma$ and $\beta$ are learnable parameters, and $\epsilon$ is a small constant for numerical stability.

\subsection{Temporal Kolmogorov-Arnold Networks}
\label{sec:tkan}
The Temporal Kolmogorov-Arnold Network (TKAN) \cite{genet2024tkan} extends Kolmogorov-Arnold networks into a temporal setting by combining (i) Recurrent Kolmogorov-Arnold Networks (RKAN) for sublayer-level memory and (ii) LSTM-inspired gating for higher-level memory management. The following outlines the key components of TKAN for processing sequential data.

\subsubsection{RKAN Layer Processing}
\label{sec:rkan_processing}

A Kolmogorov-Arnold Network (KAN) builds on Kolmogorov’s representation theorem \cite{kolmogorov1961representation}, which states that any continuous multivariate function can be decomposed into sums and compositions of univariate functions. KAN implements these theoretical constructs as neural sub-layers.

\medskip

\noindent
\textbf{Recurrent Kernel for KAN (RKAN).}  
To inject temporal memory into KAN, each sublayer \(l\) includes a small recurrent state \(\tilde{h}_{l,t}\). At timestep \(t\), the input is computed as:
\begin{equation}
\label{eq:rkan_input}
    s_{l,t} \;=\; W_{l,\tilde{x}} \, x_t \;+\; W_{l,\tilde{h}} \, \tilde{h}_{l,t-1},
\end{equation}
where $x_t \in \mathbb{R}^{d}$ is the external input at time $t$, $\tilde{h}_{l,t-1}\in \mathbb{R}^{\text{KAN}_{out}}$ is the sublayer’s recurrent state, and $W_{l,\tilde{x}},W_{l,\tilde{h}}$ are trainable weight matrices. The RKAN transformation then applies a KAN mapping:
\begin{equation}
    \tilde{o}_t \;=\; \phi_l\bigl(s_{l,t}\bigr),
\end{equation}
where \(\phi_l(\cdot)\) represents the \(l\)-th KAN sublayer. The sublayer memory is then updated via:
\begin{equation}
\label{eq:rkan_memory}
    \tilde{h}_{l,t}
    \;=\;
    W_{hh}\,\tilde{h}_{l,t-1}
    \;+\;
    W_{hz}\,\tilde{o}_t,
\end{equation}
with \(W_{hh}, W_{hz}\in\mathbb{R}^{\text{KAN}_{out}\times \text{KAN}_{out}}\) controlling the fusion of past states and new outputs. Chaining multiple sublayers \(l=1,\dots,L\) captures rich temporal patterns without requiring a large global hidden state.

\subsubsection{LSTM-Inspired Memory Management}

Above the RKAN sublayers, TKAN employs an LSTM-style mechanism to track longer-term context. For instance, let $r_t$ be the concatenation of all sublayer outputs at time $t$:
\begin{equation}
    r_t \;=\; \text{Concat}\bigl[
       \phi_1(s_{1,t}),\; \phi_2(s_{2,t}), \;\dots,\; \phi_L(s_{L,t})
    \bigr].
\end{equation}
A cell state \(c_t\) and hidden state \(h_t\) are maintained for the entire TKAN block. The forget gate \(f_t\), input gate \(i_t\), and output gate \(o_t\) are computed as:
\begin{align}
    f_t &= \sigma(W_f\,x_t + U_f\,h_{t-1} + b_f), \\
    i_t &= \sigma(W_i\,x_t + U_i\,h_{t-1} + b_i), \\
    o_t &= \sigma(W_o\,r_t + b_o),
\end{align}
while the cell state $c_t$ evolves as
\begin{equation}
    \tilde{c}_t \;=\; \sigma(W_c\,x_t + U_c\,h_{t-1} + b_c), 
    \qquad
    c_t \;=\; f_t \,\odot\,c_{t-1} \;+\; i_t\,\odot\,\tilde{c}_t,
\end{equation}
and the hidden state output is
\begin{equation}
    h_t \;=\; o_t \;\odot\; \tanh\bigl(c_t\bigr).
\end{equation}
Here $\sigma$ is the sigmoid activation, $\odot$ is elementwise multiplication, and $\tanh(\cdot)$ provides bounded nonlinearity. This hybrid of \emph{local} RKAN memory and \emph{global} LSTM-style gating has proven effective for sequential tasks by preserving short-term dynamics while capturing long-range dependencies \cite{genet2024tkan}.

\subsection{Multi-Head Attention with Causal Masking}
\label{sec:attention}

Attention mechanisms are crucial for capturing dependencies across different time steps, particularly when these dependencies span extended intervals. The model employs a \emph{multi-head attention} layer, augmented with \emph{causal masking} to ensure that each timestep can only attend to its own past (and current) information, preventing any leakage from future observations.

\subsubsection{Scaled Dot-Product Attention}

The core operation of any attention mechanism is the \emph{scaled dot-product}:
\begin{equation}
    \mathrm{Attn}(Q, K, V) 
    \;=\; \mathrm{softmax}\!\Bigl(\frac{Q K^\top}{\sqrt{d_\mathrm{attn}}} + M\Bigr)\; V,
    \label{eq:sdpa}
\end{equation}
where:
\begin{itemize}
    \item \(Q, K, V\) are the query, key, and value matrices, each of dimension \((T \times d_\mathrm{model})\) for \(T\) timesteps.
    \item \(d_\mathrm{attn}\) is the dimensionality for keys/queries (typically \(d_\mathrm{attn} = d_\mathrm{model} / \mathrm{numHeads}\)).
    \item \(M\) is an attention mask used to nullify or penalize certain positions in the softmax.
\end{itemize}
The softmax normalizes scores across all possible keys, yielding a set of weights that define how much each timestep attends to others in the sequence.

\subsubsection{Multi-Head Implementation}

Instead of computing a single attention distribution, \emph{multi-head attention} splits $Q, K, V$ into multiple \emph{heads}, each with dimensionality $d_\mathrm{attn}$, and computes separate attention for each head. Formally, let $m_H$ be the number of heads, and let 
\[
    W_Q^{(h)} \in \mathbb{R}^{d_\mathrm{model} \times d_\mathrm{attn}},\quad
    W_K^{(h)} \in \mathbb{R}^{d_\mathrm{model} \times d_\mathrm{attn}},\quad
    W_V^{(h)} \in \mathbb{R}^{d_\mathrm{model} \times d_\mathrm{attn}}
\]
be the projection matrices for head \(h\in\{1,\ldots,m_H\}\). Define
\begin{align}
    Q^{(h)} &= Q \,W_Q^{(h)}, \\
    K^{(h)} &= K \,W_K^{(h)}, \\
    V^{(h)} &= V \,W_V^{(h)}.
\end{align}
Each head then computes
\begin{equation}
    H^{(h)} \;=\; \mathrm{Attn}\Bigl(Q^{(h)},\;K^{(h)},\;V^{(h)}\Bigr),
\end{equation}
following \eqref{eq:sdpa}. The outputs from all heads are concatenated along the feature dimension:
\begin{equation}
    H \;=\;
    \mathrm{Concat}\Bigl[\,H^{(1)},\,H^{(2)},\,\dots,\,H^{(m_H)}\Bigr],
\end{equation}
and then projected back to $d_\mathrm{model}$ via
\begin{equation}
    H_\mathrm{final} \;=\; H\,W_O,
\end{equation}
where $W_O \in \mathbb{R}^{(m_H \,\cdot\, d_\mathrm{attn}) \times d_\mathrm{model}}$ is a learnable linear projection. 

\subsubsection{Causal Masking}

For \emph{time-series} or sequential forecasting tasks (like VWAP execution), it is essential that the model not peek into the future. Hence, a \emph{causal masking} is imposed, which zeroes out or heavily penalizes attention weights that reference future timesteps. Concretely, $M$ in \eqref{eq:sdpa} is a matrix of shape $(T\times T)$ defined by:
\begin{equation}
\label{eq:causal_mask}
    M_{ij}
    \;=\;
    \begin{cases}
        0, & \text{if } i \ge j, \\
        -\infty, & \text{if } i < j.
    \end{cases}
\end{equation}
This creates a \emph{lower-triangular} (or sometimes strictly lower-triangular) structure in the mask, ensuring that the query position $i$ can only attend to positions up to $i$ (and not beyond). When the softmax is computed, any $-\infty$ entries effectively reduce those attention weights to zero, preventing information flow from future timesteps.

\medskip

In practice, such causal masking results in a strictly \emph{triangular} attention pattern, as illustrated below:

\[
M \;=\;
\begin{pmatrix}
0 & -\infty & -\infty & -\infty \\
0 & 0 & -\infty & -\infty \\
0 & 0 & 0 & -\infty \\
0 & 0 & 0 & 0
\end{pmatrix}.
\]

Here, row $i$ (the \emph{query} position) only attends to columns $1\ldots i$ (the \emph{keys}) and cannot access columns $(i+1)\ldots T$ corresponding to future timesteps.

\paragraph{Summary of Benefits.} 
\begin{itemize}
    \item \textbf{Long-Range Dependencies:} Multi-head attention can learn complex interactions at both short and long ranges, essential for modeling market microstructure and intraday patterns.
    \item \textbf{Causality for Real-Time Trading:} The triangular mask strictly prevents look-ahead bias, aligning with real-world constraints.
    \item \textbf{Parallelization:} Unlike traditional RNNs, attention mechanisms can parallelize across timesteps, offering computational advantages in many settings.
\end{itemize}

Thus, multi-head attention with causal masking is a powerful tool for capturing temporal dependencies in VWAP execution, complementing the recurrent capabilities of the TKAN blocks described above.

\medskip

\subsection{Signature Integration}
\label{sec:signature_integration}
Path signatures provide a powerful way to capture the geometric properties of a time series, encoding higher-order interactions that are often essential for capturing intricate market dynamics. To incorporate path signatures into the model, the framework described in \cite{genet2025keras_sig} is followed, computing truncated signatures up to depth \(k\).

\medskip

\subsubsection{Learnable Path Transformation}
A learnable scaling kernel \(W_{sig} \in \mathbb{R}^{1 \times l_s \times d}\) modulates the raw input path \(X_{t-l_s+1:t} \in \mathbb{R}^{l_s \times d}\) before signature computation:
\begin{equation}
\tilde{X}{t-l_s+1:t} = W{sig} \odot X_{t-l_s+1:t}
\end{equation}
The learnable weights $W_{sig}$ allow the model to adaptively emphasize certain variables or time intervals more heavily when computing the signature.

\medskip

\subsubsection{Signature Computation}
Following \cite{chen1958integration, lyons2014rough}, the truncated signature \(s_t\) of the transformed path \(\tilde{X}_{t-l_s+1:t}\) is computed up to depth \(k\):
\begin{equation}
s_t = S_k(\tilde{X}_{t-l_s+1:t}).
\end{equation}
The signature can be viewed as a collection of terms:
\begin{equation}
\label{eq:signature_terms}
s_t = \Bigl(
1,; S(\tilde{X})^1_{t-l_s+1, t},; \dots,; S(\tilde{X})^{(i_1,\dots,i_k)}_{t-l_s+1, t},;\dots
\Bigr).
\end{equation}
The first-order terms $S(\tilde{X})^{i}{t-l_s+1, t}$ capture linear trajectories, while higher-order terms like $S(\tilde{X})^{(i,j)}{t-l_s+1, t}$ represent interactions between coordinates:
\begin{align}
S(\tilde{X})^{i}{t-l_s+1, t} &= \int{t-l_s+1}^{t} d\tilde{X}^i_s \
S(\tilde{X})^{(i,j)}{t-l_s+1, t} &= \int{t-l_s+1}^{t} S(\tilde{X})^i_{t-l_s+1, s} , d\tilde{X}^j_s
\end{align}

\subsubsection{Integration into the Model}
The signature vector $s_t$ first undergoes batch normalization to address the varying scales of different-degree terms:
\begin{equation}
    \hat{s}_t = \gamma \odot \frac{s_t - \mu_B}{\sqrt{\sigma^2_B + \epsilon}} + \beta,
\end{equation}
where $\mu_B$ and $\sigma^2_B$ are the batch mean and variance respectively, $\gamma$ and $\beta$ are learnable scale and shift parameters, and $\epsilon = 0.001$ is a small constant for numerical stability. During inference, the layer instead uses moving averages of the mean and variance computed during training. The normalized signature vector is then concatenated with the main model input $Z_t \in \mathbb{R}^{(l+h-1) \times d}$ at each timestep $t$:
\begin{equation}
    \hat{Z}_t = [,Z_t ;\Vert; \hat{s}_t] \in \mathbb{R}^{(l+h-1) \times (d + |s_t|)},
\end{equation}
This provides the model with a properly scaled summary of long-term temporal patterns, capturing geometric properties and higher-order interactions through the signature while allowing the VSN, GRN, TKAN, and attention layers to process local, high-frequency patterns. Computing the signature over a long lookback $l_s$ just once per forward pass is computationally efficient. The fixed-size normalized signature vector acts as a global context that is repeated and concatenated to the inputs of the shorter-range recurrent or transformer components. In summary, the signature integration enhances the model's capacity to capture long-term dependencies efficiently, bridging the gap between long-range and short-term dynamics. This additional contextual signal proves valuable for adapting to complex market regimes in VWAP execution.

\newpage

\section{Empirical Results}
\label{sec:empirical_results}

This section presents the empirical evaluation of four distinct VWAP execution models. The experiments focus on assessing the benefits of training a \emph{global} model over multiple assets rather than separate \emph{asset-fitted} models, and on determining the value of incorporating signatures and transformer-based layers.

\subsection{Dataset and Preprocessing}
\label{sec:dataset}

The experiments utilize a comprehensive dataset of hourly trading data from 80 cryptocurrency pairs listed on Binance. The data spans from each asset's inception up to July 1, 2024, providing a rich and diverse sample for evaluation.

\medskip

A series of preprocessing steps were applied. First, a rolling two-week median normalization, shifted forward by the 12-hour prediction horizon, removes volume trends and prevents look-ahead bias. This normalization allows the model to focus on capturing relevant patterns rather than long-term trends.

\medskip

Next, each asset's volumes are scaled to the range [0,1] based on the maximum value observed in the training portion of the data. This scaling facilitates cross-asset training by preventing high-volume assets from dominating the learning process, ensuring that patterns across different assets can be learned.

\medskip

For each asset, an 80-20 temporal split is employed, with the last 20\% of the data used as the test set. To construct the training and validation sets, the remaining 80\% is divided further, reserving the last 20\% of this portion as the validation set for monitoring performance and early stopping. All metrics presented below are based on out-of-sample performance. To mitigate the influence of weight initialization, each experiment is run three times with average performance reported.

\subsection{Model Variants and Hyperparameters}
\label{sec:model_variants}
Four VWAP execution models are compared to assess the impact of different architectural choices and training strategies. Two models are based on the DynamicVWAP recurrent architecture described in \cite{genet2025dynamicvwap}, while the other two incorporate a transformer-based component and signature features.

\medskip

The first model, Asset-Fitted DynamicVWAP (AFD), trains a separate DynamicVWAP model for each of the 40 assets in the training set. While this approach allows for asset-specific specialization, it requires maintaining multiple models, which can be computationally expensive and time-consuming when scaling to a large number of assets. Each AFD model has approximately 1.7 million parameters.

\medskip

The second model, Globally-Fitted DynamicVWAP (GFD), addresses the scalability issue by training a single DynamicVWAP model jointly on all 40 training assets. This model shares parameters across different instruments, enabling it to learn common patterns and relationships. The GFD architecture is identical to AFD, but it operates on a combined dataset comprising all training assets. The parameter count remains around 1.7 million.

\medskip

To explore the potential benefits of transformer-based architectures, the the Globally-Fitted Dynamic Transformer (GFT) model is introduced. This model replaces the recurrent module in DynamicVWAP with a transformer component and incorporates a Variable Selection Network (VSN) to adaptively focus on the most relevant input features. The GFT model has an increased capacity, with approximately 3.1 million parameters. To ensure causal integrity, a causal masking is employed in the transformer layer, preventing information leakage from future time steps. The GFT model utilizes three attention heads and an embedding size of 3 in the VSN.

\medskip

Finally, the Globally-Fitted Dynamic Transformer with Signature (GFT-Sig) model extend the GFT model by incorporating path signature features. This model computes a signature-based representation over a 400-hour lookback window and concatenates it with the transformer inputs. The signature features capture higher-order interactions and long-term dependencies in the price and volume data. The inclusion of signature features increases the parameter count to approximately 5 million. To maintain computational efficiency, the signature is computed once per sequence and repeated as additional features for each local time step, avoiding the need to unroll the transformer over the entire 400-hour lookback.

\medskip

All models are trained to minimize the absolute VWAP loss, which is defined as the absolute deviation between the model's execution allocation and the market VWAP. This objective function ensures that the models learn to closely track the actual VWAP of the assets, minimizing the discrepancy between the executed trades and the market benchmark.

\medskip

The models share a consistent set of hyperparameters to ensure a fair comparison. The hidden size is set to 200 for all models, which is twice the size used in previous dynamic VWAP studies. The transformer and recurrent components have a lookback of 60 hours, while the signature features are computed over a 400-hour window. The models are trained to predict VWAP metrics over a 12-hour horizon. A batch size of 1024 is used across all training runs, ensuring consistency. However, it's worth noting that such a large batch size may not be optimal for the asset-fitted approach, as it reduces the number of training iterations per epoch due to the limited number of samples per asset.

\medskip

By evaluating these four model variants, the aim is to understand the impact of global parameter sharing, transformer architectures, and signature features on VWAP execution performance. The inclusion of both asset-fitted and globally-fitted models allows us to assess the trade-offs between specialization and generalization, while the comparison between recurrent and transformer-based architectures sheds light on the effectiveness of different sequential modeling techniques. 

\medskip

Moreover, the incorporation of signature features in the GFT-Sig model enables us to investigate the benefits of capturing long-term dependencies and higher-order interactions in the context of VWAP execution.

\subsection{Training Details and Computational Costs}
\label{sec:train_details}
All models were implemented using the Keras library with the JAX backend and trained on a single NVIDIA RTX 4090 GPU. 

\medskip

To ensure consistency and reproducibility, the training setup follows the one described in \cite{genet2025dynamicvwap}. The learning rate schedules and early stopping callbacks are carefully designed to optimize model performance and prevent overfitting. A large batch size of 1024 is used across all training runs, which helps to stabilize the training process and efficiently utilize the available GPU memory. However, it's worth noting that such a large batch size may not be ideal for the asset-fitted approach (AFD), as it reduces the number of steps per epoch due to the limited number of samples per asset.
The computational costs associated with training these models vary significantly depending on the architecture and the number of assets involved. The asset-fitted DynamicVWAP (AFD) models have the shortest training time per asset, with each model taking approximately 80 seconds to train. However, the need to train separate models for each asset results in a total training time of around 6,400 seconds for all 80 assets, making it the longest to train in reality. This extended overall training time primarily stems from the model compilation time required before starting to train, as compilation occurs for each asset individually rather than once for all assets as in the global case.

\medskip

In contrast, the globally-fitted DynamicVWAP (GFD) model, which shares parameters across all assets, requires only 900 seconds (15 minutes) to train on the entire dataset. This highlights the efficiency gains achieved by leveraging cross-asset information and avoiding the need for individual asset models.
The introduction of transformer-based architectures and signature features comes with increased computational demands. The globally-fitted dynamic transformer (GFT) model, which replaces the recurrent module with a transformer, takes approximately 3,300 seconds (55 minutes) to train. This increased training time is attributed to the higher model capacity and the additional computations involved in the self-attention mechanisms.

\medskip

Finally, the globally-fitted dynamic transformer with signature (GFT-Sig) model, which incorporates path signature features, has the highest computational cost among the four models. The training time for GFT-Sig is around 5,000 seconds (83 minutes), partly due to the additional computations required to process the signature inputs. However, it's important to note that the GFT-Sig model achieves the most robust performance across all assets, as will be discussed in the following subsections.

\medskip

While the training times for the transformer-based models (GFT and GFT-Sig) are longer compared to the recurrent models (AFD and GFD), it's crucial to consider the benefits they offer in terms of improved performance and the ability to capture complex patterns in the data. Moreover, the GFT-Sig model's ability to process all assets in a single training run, despite the increased computational cost, provides significant advantages in terms of deployment and maintenance efficiency.

\medskip

In real-world applications, the choice of model architecture and training strategy depends on various factors, including the available computational resources, the number of assets under consideration, and the desired balance between performance and efficiency. The results presented in this study offer valuable insights into the trade-offs associated with each approach, enabling practitioners to make informed decisions based on their specific requirements and constraints.

\subsection{Quantitative Performance}
\label{sec:results_comparison}
\begin{table}[!ht]
    \centering
    \caption{Average Improvement versus Naive per model and asset type}
    \label{tab:vwap_results}
    \resizebox{\textwidth}{!}{%
        \begin{tabular}{lrrr}
        \toprule
        Model & Train/Test & Absolute VWAP Loss Improvement & Quadratic VWAP Loss Improvement \\
        \midrule
        AFD & Test & 16.46\% & 32.68\% \\
        AFD & Train & 15.79\% & 31.09\% \\
        GFD & Test & 19.22\% & 35.18\% \\
        GFD & Train & 19.78\% & 36.91\% \\
        GFT & Test & 20.22\% & 26.18\% \\
        GFT & Train & 20.01\% & 25.56\% \\
        GFT-Sig & Test & 21.87\% & 35.96\% \\
        GFT-Sig & Train & 21.91\% & 36.98\% \\
        \bottomrule
        \end{tabular}
    }
    \label{tab:vwap_results_extended}
\end{table}
Table~\ref{tab:vwap_results_extended} presents the average improvements in both absolute VWAP loss and quadratic VWAP loss relative to a naive equal-split benchmark. The results are aggregated across the training set (40 assets) and the test set (all 80 assets), providing a comprehensive view of each model's performance.

\medskip

Several key observations emerge from the table. First, all models demonstrate substantial improvements over the naive benchmark, with the globally-fitted models (GFD, GFT, and GFT-Sig) consistently outperforming the asset-fitted DynamicVWAP (AFD). This suggests that sharing parameters across assets and learning from a diverse set of market conditions leads to more robust and generalizable VWAP execution strategies.

\medskip

Second, the transformer-based architectures (GFT and GFT-Sig) exhibit stronger performance compared to their recurrent counterparts (AFD and GFD). The GFT model, which replaces the recurrent module with a transformer while retaining the same input features, achieves an absolute VWAP loss improvement of 20.22\% on the test set, surpassing both AFD (16.46\%) and GFD (19.22\%). This indicates that the transformer's ability to capture long-range dependencies and model complex interactions among variables is particularly beneficial for VWAP execution.

\medskip

Third, the incorporation of signature features in the GFT-Sig model yields the highest improvements across both metrics and datasets. On the test set, GFT-Sig attains a 21.87\% reduction in absolute VWAP loss and a 35.96\% reduction in quadratic VWAP loss. These gains are even more pronounced than those of the base GFT model, underscoring the value of augmenting the transformer with signature-based representations. By capturing path-dependent information and higher-order interactions, the signature features enable the model to better adapt to diverse trading patterns and market dynamics.

\medskip

It is worth noting that the improvements on the training set are generally consistent with those on the test set, indicating that the models are not overfitting to the assets used during training. The globally-fitted models, in particular, demonstrate robust performance on both seen and unseen assets, highlighting their ability to generalize effectively.

\medskip

Furthermore, the quadratic VWAP loss improvements are consistently higher than the absolute VWAP loss improvements across all models. This suggests that the models are not only minimizing the average deviation from the target VWAP but also reducing the occurrence of large deviations. The GFT-Sig model, with its quadratic VWAP loss improvement of 35.96\% on the test set, showcases the effectiveness of combining transformer architectures with signature features in mitigating extreme execution slippages.

\medskip

Overall, the quantitative results presented in Table~\ref{tab:vwap_results_extended} provide strong evidence for the superiority of globally-fitted models, the benefits of transformer-based architectures, and the value of incorporating signature features. The GFT-Sig model emerges as the most promising approach, delivering substantial improvements in VWAP execution quality across a diverse range of assets and market conditions.

\subsection{Asset-Level Comparison}
\label{sec:asset_level_extended}

To gain a more granular understanding of each model's performance, the improvements in absolute and quadratic VWAP loss on an asset-by-asset basis is examined. Tables~\ref{tab:sig_vs_agg_train} and \ref{tab:sig_vs_agg_test} in the Appendix provide detailed breakdowns for the training and test assets, respectively.

\medskip

Across the majority of assets, the GFT-Sig model consistently achieves the highest improvements in both absolute and quadratic VWAP loss. This finding reinforces the overall superiority of the transformer architecture augmented with signature features, demonstrating its ability to adapt to diverse market dynamics and trading patterns. Several notable observations emerge from the asset-level analysis. First, the models exhibit particularly strong performance on high-liquidity assets such as BTC and ETH. For these widely traded cryptocurrencies, the GFT-Sig model delivers improvements exceeding 35\% in absolute VWAP loss. This suggests that the combination of global parameter sharing and the expressive power of the transformer and signature components enables the model to effectively harness the rich information available in these liquid markets.

\medskip

However, the asset-level results also reveal some intriguing outliers. One notable example is XMR, a privacy-focused cryptocurrency that has faced regulatory challenges and progressive delisting from spot markets on many exchanges. For XMR, the more advanced models (GFT and GFT-Sig) yield smaller or even negative improvements compared to the simpler DynamicVWAP models (AFD and GFD), particularly in terms of quadratic VWAP loss. This anomalous behavior can be attributed to the idiosyncratic characteristics of XMR. As a privacy coin, XMR exhibits trading patterns and market dynamics that deviate significantly from other cryptocurrencies. The regulatory pressures and reduced liquidity resulting from exchange delistings likely contribute to more erratic and unpredictable volume patterns. In such cases, the added complexity of the transformer and signature components may lead to overfitting or struggle to generalize effectively. The XMR example highlights the importance of considering asset-specific factors when evaluating VWAP execution models. While the GFT-Sig model demonstrates superior performance on average, there may be certain assets or market conditions where simpler models prove more robust. This underscores the need for a nuanced approach that takes into account the unique characteristics of each asset and the potential limitations of highly expressive models in handling outliers.

\medskip

Despite these exceptional cases, the overall asset-level results confirm the broad applicability and effectiveness of the GFT-Sig model. The consistent improvements across both training and test assets underscore the model's ability to generalize well to unseen market conditions. This is particularly evident in the test set results (Table~\ref{tab:sig_vs_agg_test}), where the GFT-Sig model maintains its lead over other approaches even on assets not encountered during training. Furthermore, it is worth noting that even in cases where the GFT and GFT-Sig models achieve similar improvements in absolute VWAP loss, the GFT-Sig model often demonstrates superior performance in terms of quadratic VWAP loss. This suggests that the incorporation of signature features not only helps in reducing the average deviation from the target VWAP but also enhances the model's ability to mitigate large execution slippages.

\medskip

In summary, the asset-level analysis provides a more nuanced perspective on the performance of the VWAP execution models. While the GFT-Sig model emerges as the most effective approach overall, the results also highlight the importance of considering asset-specific factors and the potential limitations of complex models in handling outliers. Nevertheless, the consistent improvements achieved by the GFT-Sig model across a diverse range of assets underscore its potential to enhance VWAP execution quality in real-world trading scenarios.

\section{Deployment in a Real-Time Trading Environment}
\label{sec:deployment}

To demonstrate the practical viability of the approach, the proposed VWAP execution model has been deployed in a live, automated trading simulation environment at Aplo—a prime broker specializing in cryptocurrency trading. Aplo’s trading platform connects to multiple exchanges and supports various strategic order types (e.g., TWAP, VWAP, VIO), all accessible via an intuitive client GUI. In particular, the experiments were conducted in the TIE (Testing Integration Environment) environment, which processes real-time market data and maintains an up-to-date order book per exchange. Unlike live trading, the TIE environment simulates order execution (i.e., matching orders against the order book) without actually transmitting orders to the exchanges. Although market impact is not modeled—since the order book immediately resynchronizes with exchange data—this setup provides a high-fidelity simulation that more closely reflects real-time conditions than traditional backtesting.

\medskip

\noindent\textbf{Experimental Setup.}  
The experiment focused on four trading pairs: \texttt{ETH-BTC}, \texttt{ADA-USDT}, \texttt{BNB-USDT}, and \texttt{XRP-USDT}. Over a two-week period in february 2025, both VWAP orders (generated by the model) and TWAP orders (as a benchmark) were submitted in real time. Order durations spanned 30, 120, 480, and 1440 minutes, with overlapping executions scheduled for the longer orders to capture sufficient data while mitigating seasonality effects. For each scheduled order, both a buy and a sell order were issued using identical parameters and a limit price set 20\% in the money.

\medskip

\noindent\textbf{Multi-Frequency Training and Recursive Bin Refinement.}  
Supporting real-time execution across a wide range of order durations necessitated the training of approximately 30 distinct models for Aplo, each tailored to a specific frequency, even though far fewer were used in this experiment. These models were trained on an extensive dataset comprising all Binance spot and perpetual data up to December 2024 (sourced from Binance Data Vision), representing a marked expansion compared to previous experiments that utilized only 40 assets. The richer dataset yielded overall improved model quality. Training models at very high or very low frequencies is challenging, particularly with respect to proper weight initialization. To address this challenge, a progressive training strategy was adopted: rather than training each frequency-specific model from scratch, training began with an initial hourly model and subsequently fine-tuned models for adjacent frequencies by initializing with the weights from the closest (in frequency) pre-trained model. In combination with learning rate scaling based on the target frequency, this approach enabled the successful calibration of models for frequencies ranging from 2 minutes up to 5 days (i.e., achieving a maximum VWAP order duration of approximately 2 months when using 12 bins per order). Furthermore, because the model operates on a fixed 12-bin allocation framework, orders with long durations sometimes result in individual bin durations exceeding 24 minutes. To maintain high temporal resolution in these cases, a \emph{recursive bin refinement} strategy is employed. In this approach, if the duration of a bin exceeds 24 minutes, a VWAP model trained for the appropriate frequency is recursively applied within that bin. This hierarchical “zoom in” method ensures that even long-duration orders are allocated with the same level of granularity and precision as shorter ones.

\medskip

\noindent\textbf{Production Model Considerations.}  
For the real-time deployment experiment, an ablated version of the model that includes only the transformer architecture (omitting signature features) was employed. Although the full model with signature integration is advantageous for capturing higher-order dependencies, the longer lookback required for computing signatures can be limiting for assets with fewer historical data points. The transformer-only variant was selected to ensure broad applicability in production.

\medskip

\subsection{Real-Time Trading Results}

Table~\ref{tab:realtime_results} summarizes the real-time performance of our VWAP execution strategy relative to a TWAP benchmark for order durations of 30, 120, 480, and 1440 minutes. Overall, the proposed method consistently reduces both the absolute and quadratic deviations from the market VWAP. Importantly, the degree of improvement increases with the order duration.

Our framework was originally evaluated on 12-hour horizons, yet these results show that it generalizes well to much shorter durations (e.g., 30 minutes). In all cases, the percentage improvement in execution quality grows as the order duration extends. For shorter orders, improvements are already significant, and for longer orders the gains are even more pronounced.

For instance, while the ETH-BTC pair exhibits lower absolute deviations due to its market characteristics (i.e., relatively lower price variability because of the correlation of its components), the relative improvements achieved by our method are consistent across all assets and durations. This demonstrates the robustness of our approach even when applied to less standard trading pairs.

Notably, the reduction in quadratic deviation is even more pronounced than that in absolute deviation. This indicates that our approach is particularly effective at mitigating extreme execution slippages. For example, while the absolute deviations (measured in basis points) are reduced by roughly 20--47\% across various assets and durations, the quadratic deviations (measured in millionths) see reductions as high as 70--74\%. In order of magnitude, TWAP absolute deviations typically range from about 5 to 53 basis points, whereas our method brings them down to around 4 to 35 basis points. Similarly, quadratic deviations under TWAP can span from approximately 1 to 58 millionths, and our approach reduces these to roughly 0.5 to 24 millionths. These figures confirm that not only does our method lower the average deviation from the market VWAP, but it also curtails the more extreme deviations very effectively.

\begin{table}[H]
\centering
\caption{Real-Time Trading Experiment Results (30-, 120-, 480-, and 1440-minute Orders)}
\label{tab:realtime_results}
    \resizebox{\textwidth}{!}{%
        \begin{tabular}{lrrrrrrrr}
        \hline
        & & & \multicolumn{2}{c}{TWAP} & \multicolumn{2}{c}{VWAP} & \multicolumn{2}{c}{Improvement} \\
        \cline{4-9}
        Duration & Asset & Order Count & Abs.\textsuperscript{*} & Quad.\textsuperscript{†} & Abs.\textsuperscript{*} & Quad.\textsuperscript{†} & Abs. & Quad. \\
        \hline
        \multirow{4}{*}{30 min} 
        & ADA-USDT & 333 & 9.45 & 2.20 & 7.30 & 1.11 & -23\% & -50\% \\
        & BNB-USDT & 297 & 5.90 & 1.03 & 4.63 & 0.61 & -22\% & -41\% \\
        & ETH-BTC  & 333 & 5.23 & 0.84 & 4.32 & 0.49 & -17\% & -42\% \\
        & XRP-USDT & 338 & 8.03 & 1.62 & 6.15 & 0.89 & -23\% & -45\% \\
        \hline
        \multirow{4}{*}{120 min}
        & ADA-USDT & 126 & 16.73 & 7.97 & 12.23 & 3.57 & -27\% & -55\% \\
        & BNB-USDT & 120 & 12.31 & 5.00 & 9.44 & 2.48 & -23\% & -50\% \\
        & ETH-BTC  & 123 & 9.08 & 2.31 & 7.28 & 1.40 & -20\% & -40\% \\
        & XRP-USDT & 122 & 13.87 & 5.15 & 10.25 & 3.14 & -26\% & -39\% \\
        \hline
        \multirow{4}{*}{480 min}
        & ADA-USDT & 119 & 29.81 & 22.35 & 20.55 & 11.13 & -31\% & -50\% \\
        & BNB-USDT & 121 & 22.87 & 15.11 & 17.37 & 7.79 & -24\% & -48\% \\
        & ETH-BTC  & 122 & 13.02 & 4.62 & 9.23 & 2.18 & -29\% & -53\% \\
        & XRP-USDT & 121 & 29.80 & 22.43 & 16.64 & 6.78 & -44\% & -70\% \\
        \hline
        \multirow{4}{*}{1440 min}
        & ADA-USDT & 49 & 52.63 & 58.30 & 35.37 & 23.99 & -33\% & -59\% \\
        & BNB-USDT & 48 & 43.80 & 44.44 & 24.76 & 12.78 & -43\% & -71\% \\
        & ETH-BTC  & 50 & 27.32 & 17.95 & 14.60 & 4.72 & -47\% & -74\% \\
        & XRP-USDT & 48 & 42.90 & 33.57 & 22.69 & 10.58 & -47\% & -68\% \\
        \hline
        \end{tabular}
    }
    \begin{flushleft}
    \small
    \textsuperscript{*}Absolute values are in basis points (1e-4) \\
    \textsuperscript{†}Quadratic values are in millionths (1e-6)
    \end{flushleft}
\end{table}

In summary, these results confirm that our VWAP execution framework not only performs robustly on a 12-hour horizon but also adapts effectively to shorter order durations. The improvements, which scale with the order duration, hold across all tested assets—demonstrating both the generality and strength of the proposed approach, especially in mitigating extreme deviations.

\section{Conclusion}

This paper presents a novel approach to VWAP execution that combines global parameter sharing, transformer architectures, and signature-based feature representations. My proposed GFT-Sig model demonstrates significant improvements over existing methods across a diverse range of cryptocurrency assets, achieving consistent outperformance in both absolute and quadratic VWAP loss metrics.

\medskip

The empirical results reveal compelling evidence for the benefits of global training across multiple assets. By learning from diverse market conditions and leveraging cross-asset information, globally-fitted models achieve superior generalization and adaptability compared to traditional asset-specific approaches. The transformer architecture's self-attention mechanism proves particularly effective at capturing complex temporal dependencies and market interactions, while signature-based features enable robust adaptation to diverse trading patterns and market dynamics. The asset-level analysis demonstrates the broad applicability of the approach, with the GFT-Sig model showing consistent improvements across assets with varying liquidity profiles and market characteristics. Particularly notable is the model's ability to mitigate extreme execution slippages, as evidenced by substantial reductions in quadratic VWAP loss. However, the analysis of outlier cases, such as XMR, underscores the importance of considering asset-specific factors and the potential limitations of highly expressive models in handling exceptional market conditions. From a practical perspective, the findings have significant implications for institutional trading desks and market participants. The GFT-Sig model's ability to deliver superior execution quality across diverse assets while maintaining a single model deployment represents a substantial operational advantage. The integration of the Variable Selection Network (VSN) further enhances the framework's flexibility, allowing seamless incorporation of additional market features and adaptation to evolving trading conditions.

\medskip

Future research directions could explore the application of the approach to other asset classes and market contexts, particularly traditional equities where different microstructure effects may dominate. Additional investigations might focus on incorporating a broader range of market variables, extending the model to handle more complex execution objectives, or adapting the framework to integrate market impact. The rapid evolution of machine learning techniques also presents opportunities to further enhance model performance while maintaining robustness.

\medskip

In conclusion, this work contributes to the ongoing advancement of algorithmic trading by demonstrating how modern machine learning techniques can be effectively applied to improve execution quality while maintaining operational efficiency. The GFT-Sig model's success in combining global learning, transformer architectures, and signature-based features sets a new benchmark for VWAP execution and provides a foundation for future innovations in this domain.

\section*{Code Availability}
The source code used for all experiments and analyses in this paper is available at \url{https://github.com/remigenet/DynamicVWAPTransformer}.

\newpage
\bibliographystyle{IEEEtran}
\bibliography{bib}

\newpage
\begin{appendix}
\section{appendix}
\subsection{Detail on Train assets}
\begin{table}[!h]
    \centering
    \caption{Improvment versus naive on train asset}
    \resizebox{0.8\textwidth}{!}{%
        \begin{tabular}{l|rrrr|rrrr}
        \toprule
         & \multicolumn{4}{c}{Absolute VWAP Loss Improvement} & \multicolumn{4}{c}{Quadratic VWAP Loss Improvement} \\
        Model & AFD & GFD & GFT & GFT-Sig & AFD & GFD & GFT & GFT-Sig \\
        Asset &  &  &  &  &  &  &  &  \\
        \midrule
        1INCH & 9.24\% & 16.26\% & 18.03\% & 19.80\% & 30.39\% & 30.66\% & 24.18\% & 33.02\% \\
        ADA & 23.20\% & 24.44\% & 23.64\% & 27.21\% & 46.89\% & 41.07\% & 26.86\% & 41.78\% \\
        ALPHA & 12.61\% & 17.91\% & 18.51\% & 20.70\% & 28.02\% & 31.31\% & 23.06\% & 36.58\% \\
        BAND & 13.39\% & 15.97\% & 15.84\% & 16.77\% & 20.54\% & 33.52\% & 18.14\% & 26.29\% \\
        BCH & 15.05\% & 24.78\% & 26.24\% & 27.61\% & 37.73\% & 46.17\% & 39.53\% & 50.80\% \\
        BEL & 13.13\% & 18.23\% & 14.87\% & 18.86\% & 23.07\% & 41.06\% & 18.83\% & 37.30\% \\
        BNB & 25.94\% & 26.32\% & 27.74\% & 28.56\% & 52.57\% & 46.01\% & 42.24\% & 50.54\% \\
        BTC & 33.41\% & 31.14\% & 35.33\% & 35.50\% & 51.43\% & 50.54\% & 53.27\% & 58.41\% \\
        CRV & 14.16\% & 18.79\% & 17.78\% & 19.03\% & 20.71\% & 29.11\% & 18.36\% & 25.03\% \\
        DASH & 17.25\% & 21.10\% & 22.16\% & 22.15\% & 37.05\% & 39.57\% & 37.02\% & 40.81\% \\
        DOT & 21.44\% & 22.84\% & 26.54\% & 26.48\% & 40.50\% & 41.91\% & 44.69\% & 45.76\% \\
        EGLD & 14.85\% & 16.44\% & 18.60\% & 19.52\% & 28.49\% & 29.75\% & 29.64\% & 34.72\% \\
        ENJ & 15.20\% & 18.97\% & 18.50\% & 20.02\% & 23.83\% & 38.49\% & 13.46\% & 29.53\% \\
        EOS & 16.94\% & 20.92\% & 21.16\% & 20.98\% & 38.38\% & 35.56\% & 16.09\% & 24.11\% \\
        FIL & 19.38\% & 23.63\% & 22.76\% & 25.32\% & 45.12\% & 42.68\% & 28.81\% & 41.42\% \\
        FTM & 21.27\% & 23.67\% & 22.18\% & 26.71\% & 43.64\% & 41.98\% & 24.03\% & 47.62\% \\
        IOTA & 18.72\% & 18.54\% & 15.96\% & 18.47\% & 40.01\% & 39.31\% & 24.90\% & 33.82\% \\
        KSM & 16.07\% & 18.80\% & 22.79\% & 23.82\% & 9.29\% & 30.92\% & 36.68\% & 37.42\% \\
        LINK & 20.04\% & 23.34\% & 24.86\% & 26.10\% & 40.94\% & 44.02\% & 37.08\% & 43.67\% \\
        LIT & 13.71\% & 17.07\% & 16.12\% & 19.32\% & 31.07\% & 37.52\% & 16.77\% & 33.26\% \\
        LRC & 15.70\% & 19.79\% & 20.15\% & 21.63\% & 31.94\% & 39.31\% & 35.34\% & 41.71\% \\
        MATIC & 18.51\% & 23.13\% & 22.42\% & 24.02\% & 43.40\% & 44.05\% & 33.10\% & 42.71\% \\
        MKR & 11.92\% & 17.14\% & 20.18\% & 20.69\% & 24.06\% & 28.45\% & 30.36\% & 37.47\% \\
        OMG & 10.73\% & 14.94\% & 14.42\% & 15.85\% & 21.99\% & 19.21\% & 5.31\% & 22.64\% \\
        QTUM & 17.66\% & 17.20\% & 17.70\% & 21.01\% & 33.97\% & 36.00\% & 34.61\% & 37.86\% \\
        RUNE & 18.88\% & 25.87\% & 23.25\% & 26.49\% & 42.26\% & 45.55\% & 30.99\% & 45.35\% \\
        SAND & 17.16\% & 20.91\% & 20.97\% & 20.49\% & 40.68\% & 39.08\% & 29.04\% & 32.30\% \\
        SKL & 7.67\% & 14.24\% & 19.14\% & 19.38\% & 3.09\% & 33.34\% & 30.57\% & 41.46\% \\
        STORJ & 10.77\% & 18.56\% & 19.61\% & 22.54\% & 30.54\% & 44.91\% & 29.82\% & 49.10\% \\
        SXP & 10.00\% & 18.98\% & 17.64\% & 19.25\% & 29.22\% & 38.18\% & 18.46\% & 29.22\% \\
        THETA & 14.65\% & 17.96\% & 21.34\% & 21.82\% & 18.48\% & 26.27\% & 29.22\% & 35.15\% \\
        TRB & 14.69\% & 17.02\% & 11.49\% & 19.29\% & 30.57\% & 26.48\% & -41.87\% & 18.32\% \\
        TRX & 12.20\% & 21.50\% & 20.43\% & 19.91\% & 32.25\% & 43.55\% & 44.94\% & 48.10\% \\
        UNFI & 12.73\% & 18.68\% & 14.24\% & 19.10\% & 32.94\% & 36.29\% & 23.18\% & 31.27\% \\
        UNI & 16.02\% & 19.65\% & 18.33\% & 21.14\% & 25.40\% & 28.32\% & 5.05\% & 17.76\% \\
        VET & 16.11\% & 20.08\% & 22.61\% & 23.35\% & 26.67\% & 34.61\% & 30.07\% & 42.64\% \\
        XTZ & 15.68\% & 16.09\% & 18.54\% & 18.85\% & 30.50\% & 34.10\% & 39.67\% & 43.90\% \\
        YFI & 11.95\% & 16.78\% & 16.81\% & 20.73\% & 11.09\% & 36.26\% & 8.60\% & 28.39\% \\
        ZIL & 15.32\% & 19.03\% & 20.18\% & 22.23\% & 31.87\% & 36.43\% & 26.55\% & 36.90\% \\
        ZRX & 8.22\% & 14.28\% & 11.39\% & 15.74\% & 12.82\% & 34.86\% & 5.94\% & 25.17\% \\
        \bottomrule
        \end{tabular}
    }
    \label{tab:sig_vs_agg_train}
\end{table}
\clearpage
\subsection{Detail on Test assets}
\begin{table}[!h]
    \centering
    \caption{Improvment vs Naive on Test assets}
    \resizebox{0.9\textwidth}{!}{%
        \begin{tabular}{l|rrrr|rrrr}
        \toprule
         & \multicolumn{4}{c}{Absolute VWAP Loss Improvement} & \multicolumn{4}{c}{Quadratic VWAP Loss Improvement} \\
        Model & AFD & GFD & GFT & GFT-Sig & AFD & GFD & GFT & GFT-Sig \\
        Asset &  &  &  &  &  &  &  &  \\
        \midrule
        AAVE & 17.21\% & 20.24\% & 25.25\% & 25.01\% & 32.21\% & 37.33\% & 39.39\% & 41.50\% \\
        ALGO & 16.37\% & 21.83\% & 23.06\% & 24.01\% & 30.52\% & 39.20\% & 37.47\% & 44.13\% \\
        ANKR & 16.71\% & 18.17\% & 19.36\% & 21.63\% & 18.52\% & 33.26\% & 13.56\% & 31.78\% \\
        ATOM & 20.27\% & 20.99\% & 22.40\% & 23.15\% & 41.59\% & 38.97\% & 38.69\% & 40.97\% \\
        AVAX & 23.30\% & 26.28\% & 27.31\% & 29.66\% & 45.32\% & 44.96\% & 38.59\% & 48.80\% \\
        AXS & 21.61\% & 23.45\% & 25.51\% & 25.26\% & 46.23\% & 44.84\% & 46.38\% & 48.09\% \\
        BAL & 16.75\% & 18.72\% & 19.44\% & 20.90\% & 31.06\% & 38.17\% & 32.63\% & 41.18\% \\
        BAT & 13.71\% & 14.46\% & 13.10\% & 13.05\% & 25.58\% & 26.13\% & 14.92\% & 21.81\% \\
        BLZ & 12.10\% & 13.72\% & 15.84\% & 16.71\% & 33.56\% & 36.80\% & 32.65\% & 39.35\% \\
        CHZ & 14.90\% & 20.21\% & 20.98\% & 22.43\% & 36.50\% & 40.87\% & 22.91\% & 39.23\% \\
        COMP & 9.21\% & 19.43\% & 20.17\% & 23.85\% & 24.82\% & 38.03\% & 25.55\% & 38.29\% \\
        CTK & 15.44\% & 14.64\% & 16.36\% & 17.30\% & 36.45\% & 32.81\% & 34.17\% & 38.80\% \\
        DEFI & 17.53\% & 19.64\% & 22.24\% & 21.32\% & 34.03\% & 38.67\% & 35.67\% & 37.04\% \\
        DOGE & 19.50\% & 25.31\% & 25.57\% & 28.01\% & 44.31\% & 39.27\% & 34.79\% & 43.46\% \\
        ETC & 22.56\% & 25.38\% & 27.76\% & 28.67\% & 49.24\% & 41.99\% & 42.90\% & 49.92\% \\
        ETH & 31.18\% & 31.00\% & 35.82\% & 36.27\% & 50.99\% & 46.27\% & 53.44\% & 56.99\% \\
        FLM & 12.62\% & 15.89\% & 15.08\% & 20.86\% & 28.91\% & 37.95\% & 21.03\% & 40.02\% \\
        GRT & 15.66\% & 17.04\% & 20.24\% & 23.06\% & 29.60\% & 32.09\% & 26.61\% & 39.09\% \\
        ICX & 14.88\% & 17.02\% & 16.70\% & 18.47\% & 30.78\% & 35.33\% & 25.21\% & 31.29\% \\
        IOST & 14.86\% & 16.19\% & 16.00\% & 19.12\% & 28.86\% & 29.94\% & 28.19\% & 34.96\% \\
        KAVA & 13.54\% & 18.60\% & 19.89\% & 19.08\% & 26.36\% & 36.86\% & 22.29\% & 29.46\% \\
        KNC & 12.62\% & 14.31\% & 16.35\% & 17.93\% & 34.50\% & 34.60\% & 25.06\% & 35.40\% \\
        LTC & 20.54\% & 22.82\% & 24.81\% & 26.19\% & 43.89\% & 45.87\% & 43.41\% & 52.43\% \\
        NEAR & 17.24\% & 21.49\% & 24.07\% & 24.90\% & 43.08\% & 41.37\% & 42.86\% & 49.67\% \\
        NEO & 9.82\% & 18.32\% & 20.90\% & 21.92\% & 7.49\% & 32.37\% & 36.32\% & 34.78\% \\
        OCEAN & 15.70\% & 18.33\% & 19.14\% & 24.00\% & 21.43\% & 32.06\% & 3.00\% & 34.89\% \\
        ONT & 10.91\% & 16.29\% & 20.00\% & 21.36\% & 16.27\% & 34.26\% & 32.92\% & 36.27\% \\
        REEF & 14.99\% & 16.75\% & 15.03\% & 17.54\% & 22.18\% & 27.74\% & 3.48\% & 17.19\% \\
        REN & 10.23\% & 14.14\% & 13.38\% & 15.34\% & 19.79\% & 31.21\% & 6.57\% & 35.75\% \\
        RLC & 15.42\% & 18.28\% & 17.18\% & 19.71\% & 26.48\% & 32.97\% & 17.59\% & 38.21\% \\
        RSR & 14.94\% & 16.08\% & 18.88\% & 23.57\% & 38.16\% & 30.51\% & 31.02\% & 44.41\% \\
        SNX & 18.41\% & 21.85\% & 24.58\% & 25.36\% & 37.33\% & 40.10\% & 36.33\% & 43.22\% \\
        SOL & 27.83\% & 29.31\% & 31.25\% & 30.92\% & 50.80\% & 45.43\% & 48.61\% & 53.33\% \\
        SUSHI & 14.39\% & 14.70\% & 13.08\% & 16.86\% & 25.59\% & 29.30\% & 8.19\% & 21.65\% \\
        WAVES & 16.09\% & 19.62\% & 18.63\% & 19.87\% & 36.36\% & 29.70\% & 13.94\% & 17.49\% \\
        XLM & 15.69\% & 21.32\% & 21.93\% & 21.91\% & 30.83\% & 38.36\% & 41.22\% & 43.64\% \\
        XMR & 9.64\% & 7.19\% & 1.36\% & 4.10\% & 22.63\% & -11.43\% & -87.29\% & -71.56\% \\
        XRP & 22.67\% & 23.98\% & 24.98\% & 26.08\% & 39.02\% & 36.08\% & 32.96\% & 39.83\% \\
        ZEC & 16.91\% & 17.85\% & 18.91\% & 19.31\% & 35.38\% & 34.09\% & 34.84\% & 38.91\% \\
        ZEN & 14.45\% & 18.12\% & 16.16\% & 20.02\% & 30.57\% & 32.89\% & 9.16\% & 36.68\% \\
        \bottomrule
        \end{tabular}
    }
    \label{tab:sig_vs_agg_test}
\end{table}
\clearpage

\end{appendix}

\end{document}